# Multimodal Transformer for Sample-Aware Prediction of Metal-Organic Framework Properties

*Seunghee Han[1‡], Jaewoong Lee[2‡], and Jihan Kim[1*]*

1 Department of Chemical and Biomolecular Engineering, Korea Advanced Institute of Science and Technology, Daejeon 34141, Republic of Korea.

2 Department of Materials Science and Engineering, Korea Advanced Institute of Science and Technology, Daejeon 34141, Republic of Korea.

*Corresponding author: jihankim@kaist.ac.kr

# ABSTRACT

Metal–organic frameworks (MOFs) are a major target of machine-learning-based property prediction, yet most models assume that a single framework representation maps to a single property value. This assumption becomes problematic for experimental MOFs, where samples reported as the same framework can exhibit different properties because of differences in crystallinity, phase purity, defects, and other sample-dependent factors. Here we introduce Experimental X-ray Diffraction Integrated Transformer (EXIT), a multimodal transformer for sample-aware prediction of MOF properties that combines MOFid with X-ray diffraction (XRD). In EXIT, MOFid encodes MOF identity, whereas XRD provides complementary information about the experimentally realized sample state. EXIT is pre-trained on one million hypothetical MOFs with simulated XRD to learn transferable representations, leading to improved downstream performance relative to existing approaches. EXIT is fine-tuned on literature-derived experimental datasets for surface area and pore volume prediction. Incorporating experimental XRD improves predictive performance relative to models without experimental XRD, and attention analysis and sample-level case studies further show that EXIT assigns different predictions to samples sharing the same MOF identity when their XRD patterns differ. These results establish a practical step from framework-aware to sample-aware MOF property prediction and highlight the value of incorporating experimental characterization into porous materials informatics.

# INTRODUCTION

Machine learning has become an important tool for predicting material properties across a wide range of systems[1], including zeolites[2,3], metal-organic frameworks (MOFs)[4,5], polymers[6,7], and inorganic solids[8]. In many of these settings, models learn a mapping from a material representation to a scalar property value and have shown strong performance when trained on large, curated datasets[9,10]. This formulation is especially natural for simulation-ready databases, where each material is represented by a well-defined structure and paired with properties computed under idealized conditions[11-13]. However, the same formulation is less straightforward for experimental materials data, where the relationship between a nominal material identity and a measured property is often not strictly one-to-one[14,15].

A central difficulty is that experimental measurements are performed not on an abstract material label, but on a realized sample. As a result, the same nominal material can exhibit different reported properties across studies because of differences in synthesis conditions, activation procedures, crystallinity, defect concentrations, and other sample-dependent factors[16-18]. In this sense, part of the discrepancy between machine-learning predictions and experimental measurements is not simply noise, but a representation problem: when the model input encodes only composition, topology, or an idealized structure, sample-level variation may be forced into the residual error rather than represented explicitly[19,20]. This issue is particularly visible in MOFs, where even widely studied frameworks such as MOF-5, HKUST-1, and UiO-66 are known to show substantial spreads in experimentally reported properties[18,21-23].

To this end, MOFs provide a useful setting in which to study this problem because their structural diversity has made them a major target of ML-based property prediction, while at the same time, their experimentally measured properties often depend strongly on sample history and preparation[24,25]. Existing ML models for MOFs typically rely on descriptors or encoded features derived from ideal crystal structures[10,26,27]. Yet such representations are fundamentally framework-level, as they encode the nominal identity of the MOF but not necessarily the state of the experimentally realized sample. For experimental MOF prediction, this creates a gap between the information available to the model and the factors that determine the reported measurement.

One practical way to reduce this mismatch is to augment framework- or identifier-based representations with an experimental descriptor that reflects sample state. Powder X-ray diffraction (XRD) is an attractive candidate for this purpose because it is routinely reported in MOF synthesis studies and is sensitive to crystallinity, phase composition, symmetry, structural distortion, and other characteristics of the realized sample that are not fully captured by nominal framework identity alone[28]. XRD is also appealing from a modelling perspective because it can be incorporated in both simulated and experimental settings. In hypothetical MOFs, diffraction patterns can be readily generated from crystal structures, enabling large-scale multimodal pre-training with paired MOF representations and XRD data. Prior studies have shown that simulated XRD patterns can support prediction of properties in inorganic crystals and porous materials[29,30]. For experimental MOFs, XRD is routinely reported as a standard characterization following synthesis, making it readily accessible for literature-based data collection.

Here we introduce EXIT (Experimental X-ray Diffraction Integrated Transformer), a multimodal framework for MOF property prediction that combines MOFid[31] with XRD information. EXIT is designed to integrate MOF-level representations with diffraction data that reflect sample-dependent characteristics. We first evaluate this framework in a controlled setting using hypothetical MOFs paired with simulated XRD patterns, and then in a more challenging experimental setting using literature-derived MOF-property-XRD datasets. Our goal here is not to claim that diffraction fully resolves the complexity of experimental variability, but rather to test whether it provides useful sample-level information beyond MOF identity alone. By treating MOF property prediction as a sample-aware problem, this work explores a practical strategy to narrow the gap between idealized MOF representations and experimentally measured behavior.

# RESULTS

## EXIT: Multimodal Architecture and Pre-training

The overall workflow of EXIT is illustrated in **Figure 1**. EXIT is a multimodal Transformer model designed to integrate two complementary inputs: MOFid and an XRD representation. MOFid encodes the idealized chemical identity of a framework in a language-like format, including the metal node, organic linker, topology, and catenation[31], whereas XRD provides complementary information on the experimentally realized crystalline state, such as phase identity, symmetry, crystallinity, crystallite domain size, strain, and preferred orientation. In the architecture used in this study, MOFid is tokenized as a sequence input, and XRD is encoded by a one-dimensional convolutional neural network before multimodal fusion.

EXIT was first pre-trained to learn universal representations of MOFs. Because large-scale paired experimental MOFid–XRD datasets are not available, we constructed a pre-training dataset by combining hypothetical MOFs generated with PORMAKE[11] and MOF structures collected from the hMOF[12], CoRE MOF[32], and QMOF databases[33]. Simulated XRD patterns for all structures were then computed using pymatgen[34]. The pre-training tasks consisted of masked language modeling (MLM) on MOFid and void-fraction prediction using the CLS token. Together, these tasks allowed the model to learn both framework-level chemical representations and global structure-related features from the XRD data. The hypothetical dataset was split into training and test sets at a ratio of 0.95:0.05, and the pre-training performance is summarized in **Table S1**.

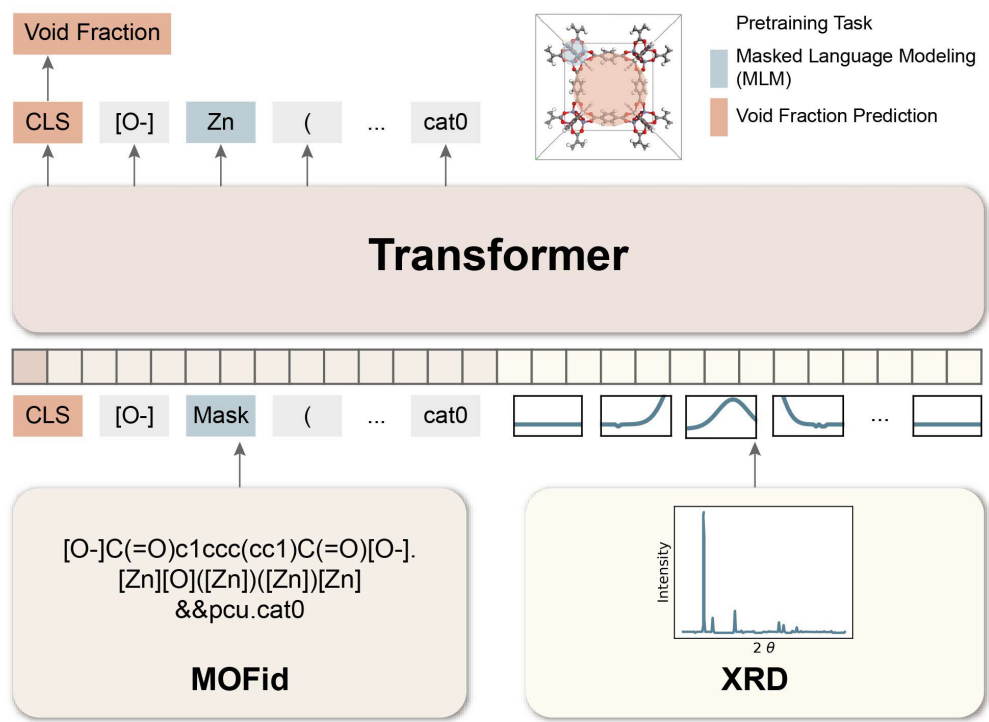

**Figure 1** Overview of the EXIT framework. MOF structures are represented using MOFid, while X-ray diffraction (XRD) patterns are incorporated as an additional modality reflecting experimentally accessible structural information. The MOFid sequence and discretized XRD signals are jointly embedded and processed by a Transformer encoder. During pre-training, the model is trained with masked language modeling (MLM) on MOFid tokens and a regression task for predicting void fraction from the [CLS] representation. This multimodal pre-training enables the model to learn representations that integrate framework identity with diffraction-derived features.

## Effect of pre-training on downstream tasks with simulated XRD

To assess the effect of pre-training and benchmark EXIT against prior models, we performed downstream prediction tasks using simulated XRD for thermal decomposition temperature ($T_D$) and $CH_4$ uptake at 298 K and 2.5 bar. For each dataset, we trained representative baselines from previous studies, including descriptor-based models for $T_D$ prediction and MOFormer, a MOFid-based model, for $CH_4$ uptake prediction. In both tasks, the pre-trained EXIT model achieved the best performance (**Table 1**). Compared with EXIT trained from scratch, pre-training substantially improved accuracy: the mean absolute error (MAE) decreased from 54.99 K to 44.58 K for $T_D$ and from 0.30 to 0.17 for $CH_4$ uptake.

We further conducted an ablation study to examine how different pre-training tasks affected downstream performance (**Table S2**). Masked language modeling contributed more strongly to $T_D$ prediction, whereas void-fraction prediction was more beneficial for CH₄ uptake prediction. Nevertheless, combining MOFid masked language modeling and void-fraction pre-training outperformed either task alone. Taken together, these results show that multimodal pre-training provides an effective initialization for downstream tasks and allows EXIT to learn transferable structure–property relationships before exposure to experimental data.

**Table 1** Effect of multimodal pre-training on property prediction using simulated XRD. Mean absolute error (MAE) is reported for thermal decomposition temperature ($T_D$) and CH₄ adsorption at 298 K and 2.5 bar. The pre-trained EXIT model consistently outperforms descriptor-based and MOFid-based baselines, as well as the same architecture trained from scratch, demonstrating the benefit of pre-training with hypothetical MOFs and simulated diffraction data.

| Mean Absolute Error (MAE) | Thermal Decomposition Temperature ($T_D$) | CH₄ Adsorption (298K, 2.5 bar) |
|---|---|---|
| Descriptor + Random Forest (RF) | 45.03 | 0.21 |
| MOFormer | 50.14 | 0.33 |
| EXIT (Scratch) | 54.99 | 0.30 |
| EXIT (pre-trained) | **44.58 ( ↓ )** | **0.17( ↓ )** |

## Construction of the Experimental Dataset

Having validated EXIT on downstream tasks with simulated XRD, we next constructed an experimental dataset to fine-tune the model for prediction of experimentally measured properties. The overall workflow of experimental dataset construction is shown in **Figure 2**.

Experimental XRD patterns were obtained using ChatMatGraph, which integrates the graph-mining tool MatGD[35] with a multimodal large language model. We started from 69,183 MOF-related papers whose DOIs were listed in the large language model–based MOF miner (L2M3) literature database reported in our previous work[24]. For each paper, figure images were downloaded from ACS and Elsevier journals using the corresponding DOI, and ChatMatGraph was used to identify figures containing XRD patterns. When a single figure contained multiple XRD panels, ChatMatGraph separated them into individual images, which were then digitized into numerical diffraction patterns. The extracted patterns were subsequently processed by baseline correction, rule-based post-processing, and min–max normalization to the range of 0 to 1. Cases in which XRD extraction failed were discarded. MOF identities were assigned primarily from graph legends through ChatMatGraph, with caption information incorporated by ChatMatGraph as supplementary context. Further details of the ChatMatGraph pipeline are provided in the Methods section.

In parallel, experimental MOFs were mapped to framework identities by checking DOI–name–refcode pairs, retrieving CIF structures through CCDC references when available, refining the structures, removing free solvent, and filtering invalid entries using MOFChecker. Additional manual refinement was applied, where possible, to correct invalid CIF files and to resolve naming ambiguities for representative MOFs such as MOF-5 and HKUST-1. MOFid was then extracted from the curated CIF structures.

Property data were collected for surface area (SA) and pore volume (PV) from L2M3[24]. For surface area, only Brunauer–Emmett–Teller (BET) values reported in m² g⁻¹ were retained. For pore volume, only total pore volume values reported in cm³ g⁻¹ were retained. Records associated with nonstandard units, value ranges, lists, or ambiguous duplicate entries were removed during curation. Despite the large initial literature corpus, the final number of usable samples was limited by the requirement that each sample include a matched MOFid, an experimental XRD pattern, and a curated property value. Many candidate entries lacked at least one of these components, for example an associated CIF structure, an extractable XRD figure, or a usable reported property value. In addition, some XRD-containing figures could not be successfully parsed by ChatMatGraph, especially when multiple diffraction panels overlapped within a single figure. The final curated dataset comprised 311 surface area records across 84 MOFs and 181 pore volume records across 49 MOFs, each paired with MOFid and experimental XRD.

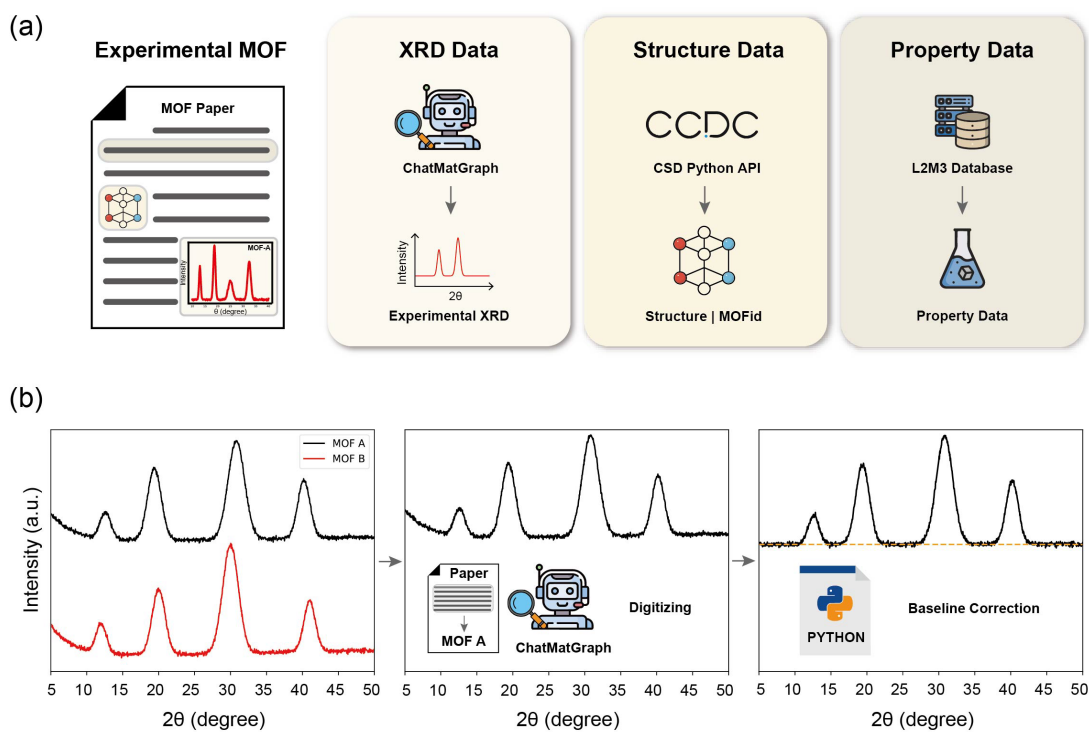

**Figure 2** Workflow for construction of the experimental MOF dataset. (a) Experimental XRD, structure, and property data were collected from the literature using ChatMatGraph, CCDC records, and the L2M3 database. (b)

XRD patterns were extracted by panel separation and digitization, followed by baseline correction.

## Experimental XRD and MOF Property Prediction

Because most previous MOF property predictors rely only on idealized framework information, they cannot distinguish samples that share the same nominal MOF identity but differ in experimentally realized state. To address this limitation, we incorporated experimentally measured XRD into EXIT to improve prediction of literature-derived experimental properties. The curated surface area and pore volume datasets were split into training, validation, and test sets at a ratio of 0.8:0.1:0.1, and the pre-trained EXIT model was fine-tuned on each task. For comparison, corresponding models were also trained without experimental XRD. For surface area prediction, incorporating experimental XRD improved the test performance from $R^2 = 0.30$ and MAE = 405 to $R^2$ = 0.53 and MAE = 334. For pore volume prediction, the model with experimental XRD also performed better, improving from $R^2 = 0.12$ and MAE = 0.26 to $R^2 = 0.59$ and MAE = 0.22 (**Figure 3**).

We further evaluated model robustness using 9-fold cross-validation. The full dataset was divided into ten folds, with one fold fixed as the test set and the remaining nine used for cross-validation. Consistent with the test-set results, models incorporating experimental XRD performed better on average than those without experimental XRD for both surface area and pore volume (**Tables S3 and S4**).

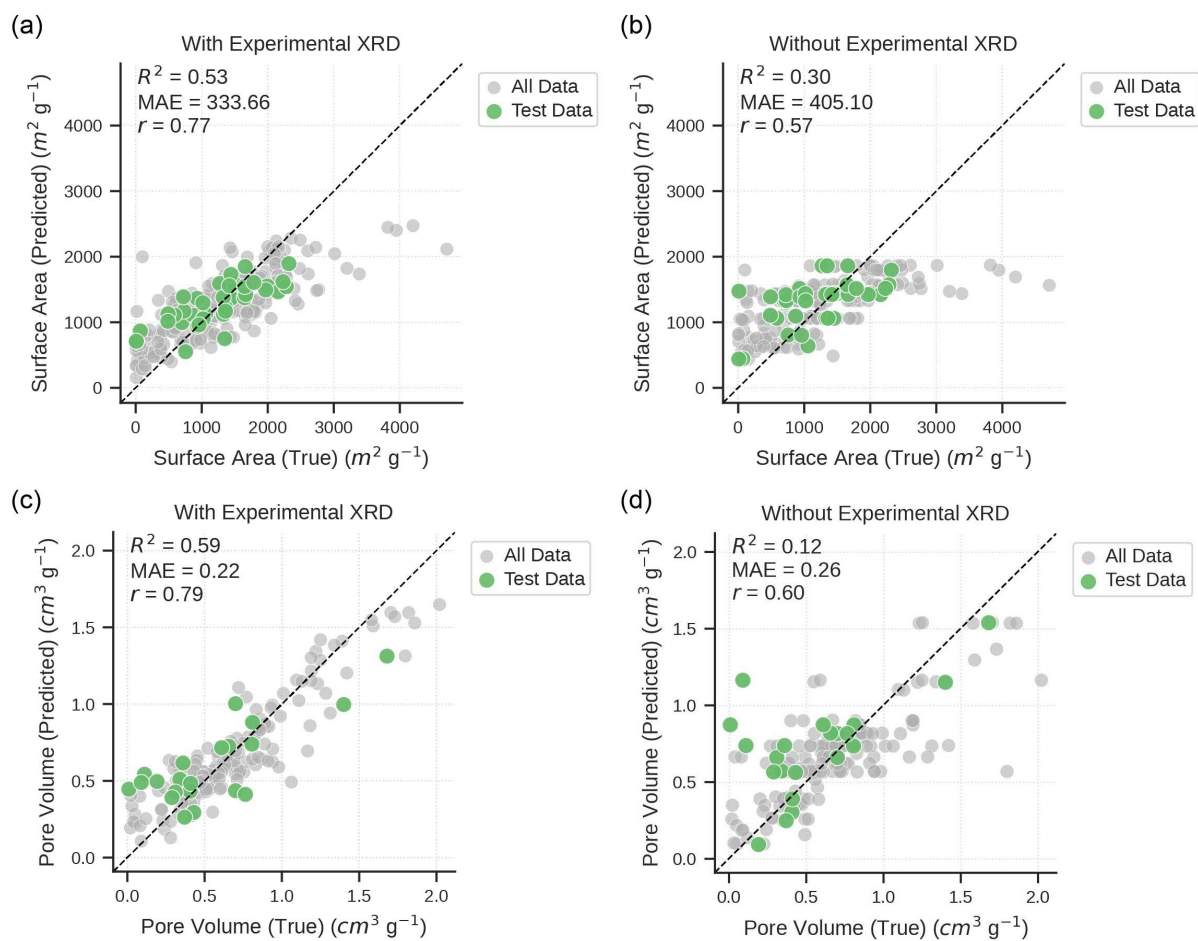

**Figure 3** Prediction results of EXIT for experimental surface area (SA) and pore volume (PV). Scatter plots compare predicted and true values for models trained (a, c) with and (b, d) without experimental XRD. Panels (a) and (b) show SA prediction, and panels (c) and (d) show PV prediction. Gray points represent all data, and green points indicate the test set. The dashed line represents the identity line. $R^2$, mean absolute error (MAE), and Pearson correlation coefficient ($r$) are reported for the test set.

This improvement is particularly meaningful because variations in surface area and pore volume can directly affect adsorption-related performance. This is also practically relevant because powder XRD is routinely collected during MOF synthesis and characterization, whereas direct measurement of SA and PV typically requires additional gas sorption experiments. As a result, XRD-informed prediction may help prioritize which experimentally synthesized samples warrant further characterization or application-specific testing. For example, $H_2$ uptake is often correlated with BET surface area, consistent with the empirical Chahine's rule, such that even samples with the same MOF may exhibit different gas adsorption or separation behavior[36].

## Attention Score Analysis

To better understand how experimental XRD enables sample-specific prediction, we examined a representative MOF-808 case together with its attention patterns (Figure 4). When experimental XRD is incorporated, EXIT assigns distinct pore-volume predictions to different MOF-808 samples. Without experimental XRD, the same samples collapse to nearly a single predicted value of approximately 0.87 $cm^3$ $g^{-1}$. This contrast suggests that, in the absence of experimental structural information, the model tends to reduce different samples of the same MOF to an averaged estimate. The attention visualizations further clarify this behavior. For the same MOF-808 samples, the MOFid attention patterns are effectively identical, whereas the XRD attention patterns differ across samples. This suggests that MOFid primarily encodes framework identity, whereas XRD captures variation associated with the experimentally realized sample state. In particular, the model appears to use differences in the presence and relative intensity of specific diffraction peaks as cues for distinguishing higher and lower pore volume.

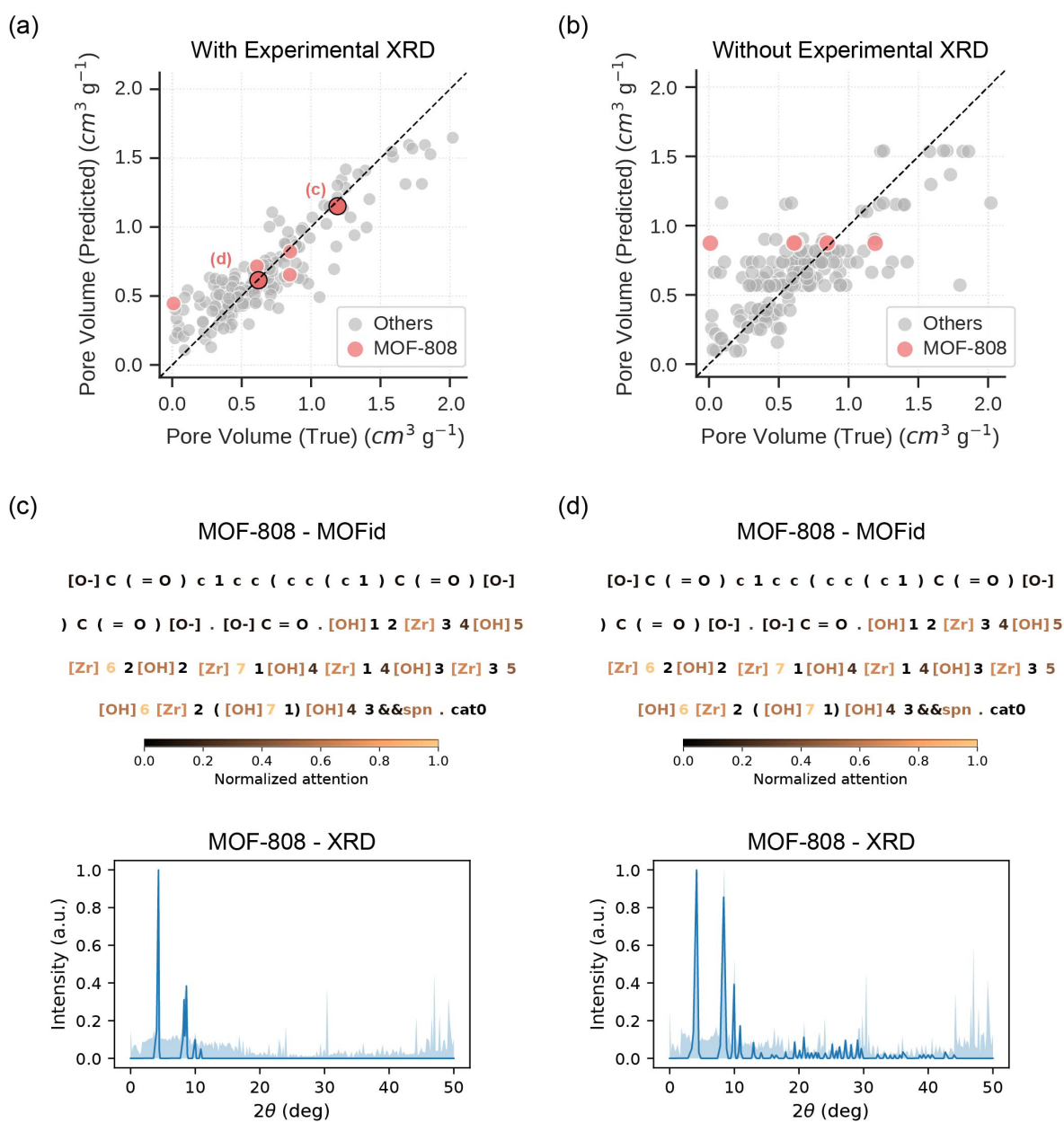

**Figure 4** Case study of MOF-808 pore volume (PV) prediction and attention analysis using EXIT. Panels (a) and (b) compare predicted and true PV values, with MOF-808 samples highlighted in red for models trained with and without experimental XRD, respectively; the dashed line in each panel indicates the identity line. Panels (c) and (d) present attention visualizations for the circled MOF-808 samples in panel (a), including attention over the MOFid tokens and the corresponding experimental XRD patterns. In the MOFid plots, attention weights are shown by color intensity, with lighter colors indicating higher attention weights. In the XRD plots, background shading indicates the attention weight.

Beyond this sample-level analysis, we next examined whether the pre-trained attention representations also capture a broader distinction between simulated and experimental XRD. For the simulated dataset, 1,000 samples were randomly selected from the pre-training test set, whereas the experimental dataset consisted of the samples used for the surface area (SA) and pore volume (PV) tasks. Attention representations extracted from the pre-trained EXIT model were projected using t-SNE, and simulated and experimental samples were clearly separated in the resulting embedding space (**Figure 5a**).

Because the combined attention representation may still be influenced by both modalities, we additionally performed t-SNE using only the XRD attention representations. In this case, the separation between simulated and experimental samples became even more distinct, with clearer clustering observed for the two groups (**Figure 5b**). These results indirectly indicate that experimental XRD differs systematically from simulated XRD. Consistent with this interpretation, a representative ZIF-8 example shows that experimental XRD can differ substantially from the corresponding simulated XRD pattern, likely due to factors such as crystallite size, defects, guest or solvent inclusion, microstrain, and activation state (**Figure 5c**).

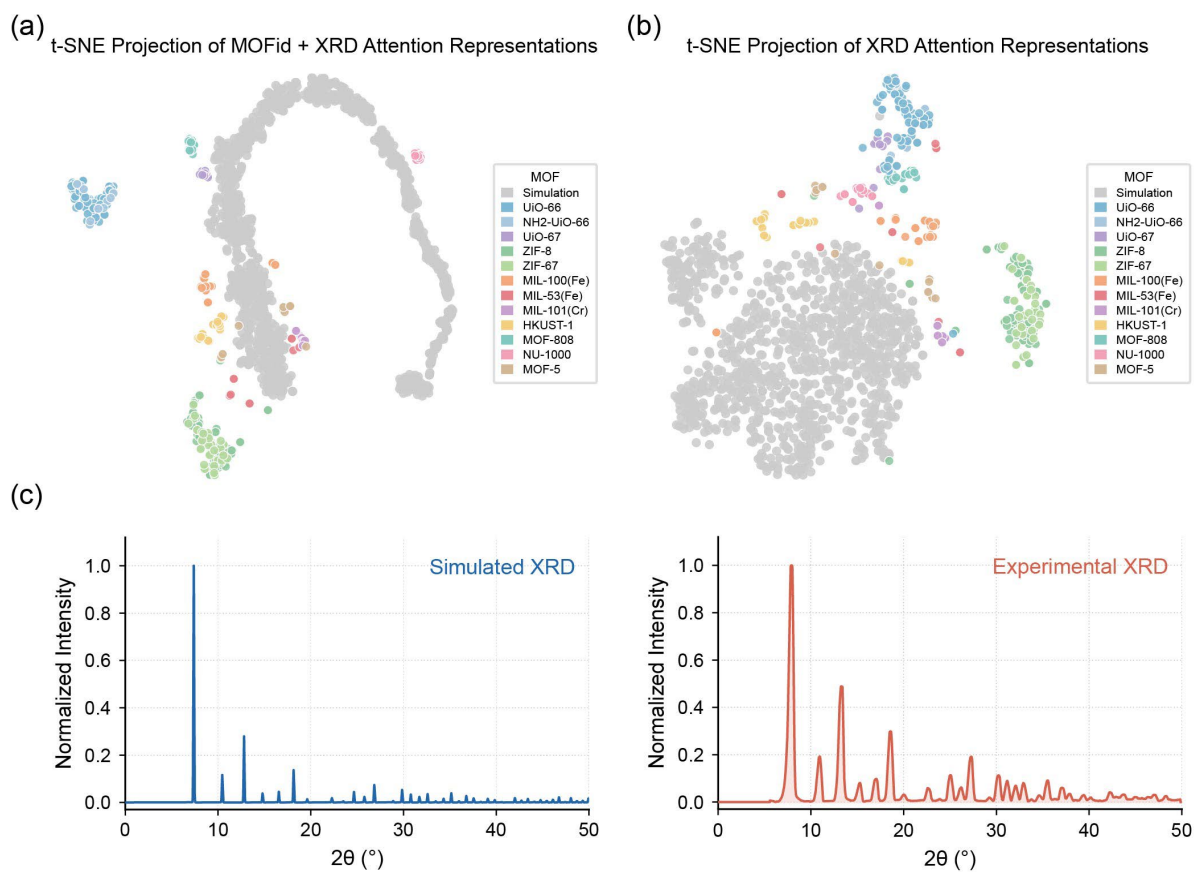

**Figure 5** t-SNE visualization of attention representations from the pre-trained EXIT model. (a) and (b) show two-dimensional projections of attention representations obtained from MOFid + XRD and XRD-only inputs, respectively. Colored points denote experimental samples from the SA and PV datasets, while gray points denote 1,000 random samples from the pre-training test set with simulated XRD. (c) compares simulated and experimental XRD patterns for ZIF-8, highlighting differences between idealized and experimentally measured diffraction signals.

## MOF-Dependent Interpretability and Limitations of EXIT

Because sample-to-sample variation in experimentally reported MOF properties may partly originate from differences in synthesis conditions, we next examined whether such metadata alone could explain the observed variation. We analyzed ZIF-8 as a representative case study because it has been synthesized using multiple Zn precursors and solvents, including Zn nitrate and Zn acetate as precursors and $H_2O$, methanol (MeOH), and DMF + MeOH as solvents. However, pore volume does not show a clear trend when grouped only by these synthesis conditions (**Figure 6**). Variation in pore volume remains even under the same precursor or solvent conditions, whereas experimental XRD improves pore volume prediction. This suggests that synthesis conditions alone are insufficient to explain sample-to-sample variation and that experimental XRD provides additional structural information relevant to pore accessibility.

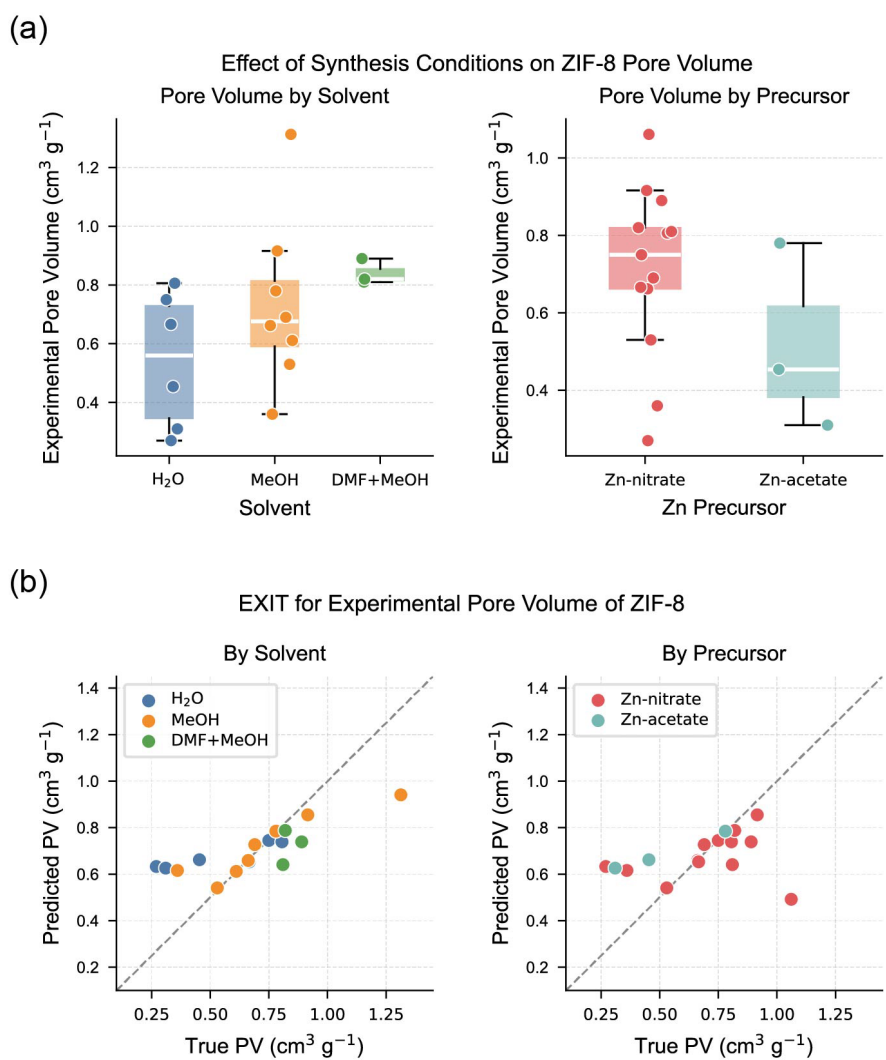

**Figure 6** ZIF-8 samples with available synthesis condition annotations for analysis of the effects of precursor and solvent on pore volume (PV). (a) Box plots show experimentally reported PV values grouped by solvent (left) and Zn precursor (right). (b) Scatter plots compare predicted and true PV values for the corresponding ZIF-8 samples, grouped by solvent (left) and Zn precursor (right).

We next examined representative cases with relatively high and limited learnability. For MOF-5, analysis suggested that the full width at half maximum (FWHM) of the dominant diffraction peak was associated with surface area. Smaller FWHM was consistent with larger coherent diffraction domain sizes and, in this dataset, with lower surface area (**Figure 7a**), although in MOFs peak width can also be affected by defects, grain boundaries, and other sources of structural disorder[37]. One possible explanation is that differences in crystalline domain structure are accompanied by differences in activation behavior, such as the efficiency of solvent removal or the degree of pore blocking, which in turn affect the accessible surface area[38,39]. Although these differences are difficult to distinguish by visual inspection of the XRD patterns alone, EXIT appears to have learned these subtle variations. While a larger Scherrer domain size does not necessarily imply a larger crystal size of the MOF itself, the result suggests that crystallinity-related features may be linked to accessible porosity[38].

In contrast, the UiO series provided representative cases with limited learnability (**Figure 7b**). In UiO-66 and UiO-67, missing-linker defects can alter properties such as surface area and pore volume. However, previous work has suggested that defect levels in UiO-66/67 are difficult to distinguish by XRD alone and instead require complementary analyses such as TGA, NMR, and EDX[40]. Consistent with this, the XRD patterns shown in **Figure 7b** (right) exhibit only minor differences despite differences in defect levels. In such cases, where the experimentally relevant variation is not clearly reflected in XRD, the predictive benefit of incorporating experimental XRD is expected to be limited.

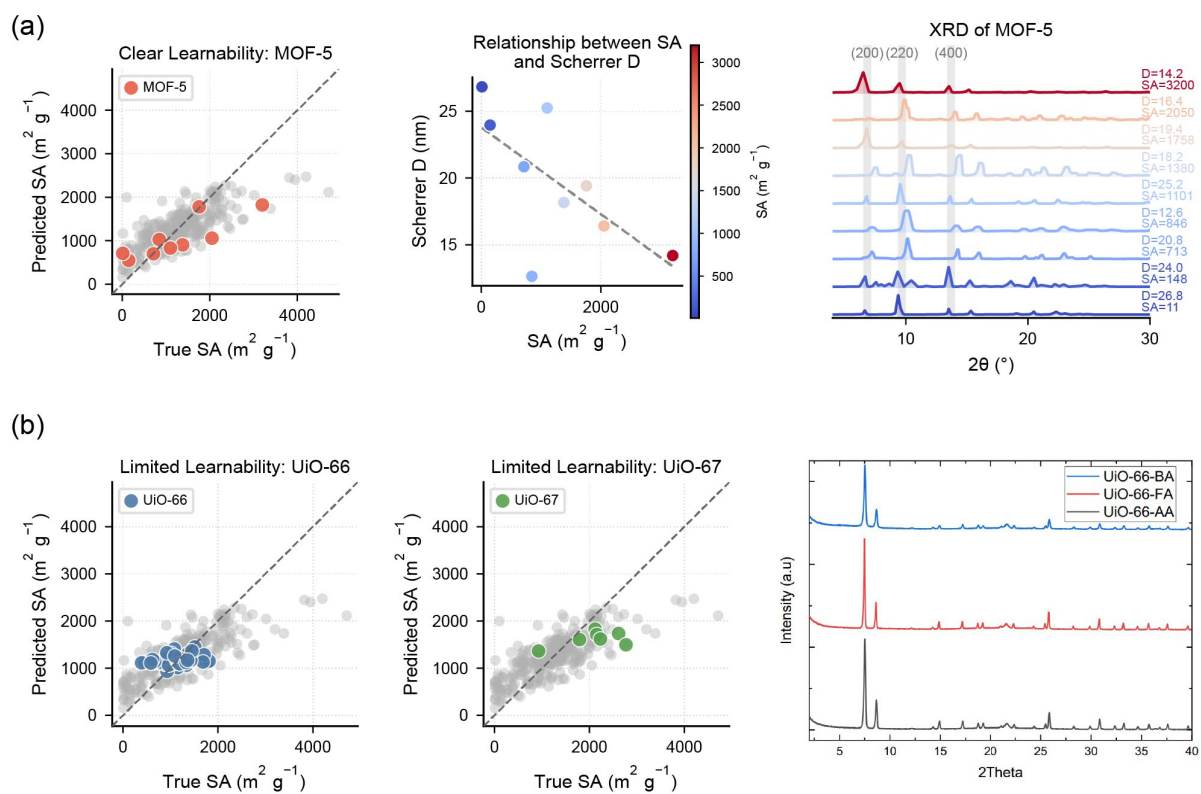

**Figure 7** Case studies illustrating MOF-dependent behavior in experimental surface area (SA) prediction with EXIT. (a) MOF-5: predicted versus true SA values (left), relationship between Scherrer domain size and SA (middle), and experimental XRD patterns for samples with different SA values (right). (b) UiO-66 and UiO-67: predicted versus true SA values for UiO-66 (left) and UiO-67 (middle), and experimental XRD patterns of UiO-66 samples with different defect levels (right). The UiO-66 XRD patterns were adapted from Ref. 40 under the CC BY 4.0 license.

Taken together, these analyses show that experimental XRD can provide sample-relevant structural information beyond idealized framework representations alone, although its usefulness depends on whether the experimentally important variation is reflected in the diffraction pattern. Finally, although EXIT requires experimental input in the form of measured XRD, this dependence remains practically attractive because powder XRD is a routine and comparatively accessible characterization method in MOF research. In many workflows, diffraction data can be acquired earlier and more readily than detailed adsorption measurements, allowing the model to use a widely available structural signal to inform sample-aware property prediction. Rather than replacing experiment, this framework suggests a way to use early-stage characterization to prioritize samples for more resource-intensive follow-up measurements.

# DISCUSSION

In this work, we introduced EXIT, a multimodal Transformer framework that integrates MOFid and XRD to enable experiment-aware prediction of MOF properties across hypothetical and experimental settings. Pre-training on one million hypothetical MOFs showed that the model can learn transferable multimodal representations from chemical identity and diffraction-derived structural information. On simulated-XRD benchmarks, the pre-trained model achieved improved downstream performance for thermal decomposition temperature and $CH_4$ uptake prediction compared with existing approaches and with the model trained from scratch. Applying EXIT to literature-derived experimental datasets further demonstrated that incorporating experimental XRD improves prediction of surface area and pore volume compared with models without experimental XRD. Attention analyses and case studies further showed that EXIT can assign different predictions to samples of the same MOF when experimental XRD is provided and that the predictive benefit of XRD depends on whether the relevant variation is reflected in the diffraction patterns. Although the present study should be viewed as a proof-of-concept rather than a definitive benchmark, given the heterogeneity of literature-derived property records and the use of XRD patterns recovered from published figures, the results nevertheless demonstrate the value of incorporating experimental characterization into MOF property learning. Taken together, these findings establish EXIT as a practical step from framework-level prediction toward sample-level prediction in porous materials informatics and highlight the broader importance of cleaner, larger, and more standardized paired experimental datasets for future data-driven materials discovery.

# METHODS

*Multimodal Transformer*

EXIT adopts a multimodal architecture that takes MOFid sequences and XRD patterns as inputs. The MOFid vocabulary contains 3,621 unique tokens, and each sequence includes up to 512 tokens. The XRD pattern is represented by normalized intensities sampled from $2\theta = 0–50°$ at a $0.01°$ interval, resulting in a 5,000-dimensional input vector. MOFid tokens are mapped to the contextual token embeddings, while the XRD signal is divided into 1D patches of size 20 and encoded into patch embeddings. The resulting MOFid and XRD token embeddings are augmented with modality-specific token-type embeddings and concatenated along the sequence dimension to form a single multimodal token sequence, which is then passed through the Transformer blocks. These multimodal Transformer blocks have a hidden dimension of 768 and consist of six layers with eight attention heads per layer.

*Pre-training*

To pre-train EXIT, we constructed a large-scale dataset by generating one million hypothetical MOF structures using PORMAKE, a Python library that builds porous materials from predefined topologies and building blocks. MOFid representations were extracted from these structures, and their corresponding XRD patterns were simulated using the pymatgen library. For the pre-training tasks, we employed masked language modeling (MLM) on the MOFid sequences and the prediction of the void fraction, which was computed using Zeo++. We used the AdamW optimizer with a learning rate of $10^{-4}$ and a weight decay of $10^{-2}$, and trained the model for 500 epochs with a batch size of 256. The dataset was randomly split into 950,000 samples for training and 50,000 samples for testing. A polynomial decay scheduler with a warm-up ratio of 0.05 was applied during training.

*Fine-tuning*

For downstream prediction of thermal decomposition temperature ($T_D$) and $CH_4$ uptake using simulated XRD, we fine-tuned the model using the AdamW optimizer with a learning rate of $1 \times 10^{-4}$ and a weight decay of $10^{-2}$. The dataset was randomly split into training, validation, and test sets at a ratio of 0.65:0.15:0.20. For prediction of surface area and pore volume using experimental XRD, the learning rate was set to $5 \times 10^{-5}$, and the dataset was split into training, validation, and test sets at a ratio of 0.8:0.1:0.1. In all cases, the model was trained for 30 epochs with a batch size of 128. A polynomial decay scheduler with a warm-up ratio of 0.05 was applied during training. For experimental surface area and pore volume prediction, we additionally performed 9-fold cross-validation by dividing the full dataset into ten folds, fixing one fold as the test set and using the remaining nine folds for cross-validation.

*ChatMatGraph*

XRD data were extracted using ChatMatGraph, a graph-mining extraction agent developed in our group by combining MatGD with GPT-5.4[41], a multimodal large language model. We first retrieved figures from papers in the L2M3 database whose DOIs were associated with reported surface area or pore volume values. ChatMatGraph then used figure captions to filter for figures containing XRD patterns. For each selected figure, the model was used to identify the correspondence between individual data lines and their associated MOF entities. The extracted XRD traces were subsequently digitized, producing experimental XRD-property pairs that connect digitized XRD patterns with MOF property values curated in the L2M3 database. To standardize the extracted signals, each digitized trace was first reduced to a one-dimensional adaptive centerline representation. Baseline correction was then evaluated using three candidate methods from the pybaselines library[42], namely arPLS, AsLS, and airPLS, and the optimal method for each trace was selected by minimizing a composite score. This score was defined as a weighted sum of three criteria: the mean magnitude of negative values in the corrected signal, the mean absolute deviation between the estimated baseline and a rolling lower-percentile envelope, and the mean absolute second derivative of the baseline, which penalized excessive roughness. The baseline-corrected traces were finally min–max normalized to the range of 0–1. All extracted XRD-property pairs were further manually inspected, and samples containing errors or ambiguities were removed during the curation process.

# Data and Code availability

The code and data are available at https://github.com/seunghhs/EXIT.git

## ASSOCIATED CONTENT

**Supporting Information**. The following contents are included: the distribution of experimentally mined MOF property data; prediction results of the best-performing EXIT model for experimental surface area and pore volume, grouped by MOF; performance of the pre-training tasks on the hypothetical MOF dataset; test-set performance on downstream fine-tuning tasks with different pre-training tasks; and cross-validation results for experimental surface area and pore volume prediction.

## AUTHOR INFORMATION

### Corresponding Author

* Email: jihankim@kaist.ac.kr

Author Contributions

    S.H. and J.L. contributed equally to this work: They conceived the idea, planned modeling, implemented the work. S.H., J.L., and J.K wrote the manuscript. All authors contributed to discussions informing the research.

### ORCID

Seunghee Han: 0000-0001-8696-6823

Jaewoong Lee: 0009-0002-8968-3292

Jihan Kim: 0000-0002-3844-8789

**Notes**

The authors declare no competing interest.

## ACKNOWLEDGEMENT

    This work was supported by the National Research Foundation of Korea (NRF) (RS-2024-00451160 and RS-2024-00435493).